\documentclass{naun}
\renewcommand{\thefigure}{\arabic{figure}}
\newcommand\blfootnote[1]{%
  \begingroup
  \renewcommand\thefootnote{}\footnote{#1}%
  \addtocounter{footnote}{-1}%
  \endgroup
}
\usepackage{subcaption}
\begin{document}

\title{The numerical solution of the free-boundary cell motility problem}

\author{Vitaly Chernik, Pavel Buklemishev \\
Ishlinsky Institute for Problems in Mechanics of the Russian Academy of Sciences,\\
Prospekt Vernadskogo 101-1, Moscow, 119526,\\
Russia}

\twocolumn[
  \begin{@twocolumnfalse}
      \date{}
      \maketitle

\hspace {1cm}\begin{minipage}{0.9\linewidth}
 
{\color{red}}
\end{minipage}

     \end{@twocolumnfalse}
]
\thispagestyle{empty}

\begin{abstract}
\indent
The cell motility problem has been investigated for a long time. Today, many biologists, physicists, and mathematicians are looking for new research instruments for this process. A simple 2D model of a free-boundary cell moving on a homogeneous isotropic surface is presented in the paper. It describes the dynamics of the complex actomyosin liquid, whose special properties influence the boundary dynamics and cell motility. The model consists of a system of equations with the free boundary domain and contains a non-local term. In this work, we present a numerical solution of this problem. 

\end{abstract}


\section{Introduction}
\lettrine{\bfseries M}{ANY} species of \blfootnote{The work was carried out on the state assignment (state registration No. AAAA-A20-120011690138-6), with the support of a grant from the Ministry of Science and Higher Education of the Russian Federation} living organisms, ranging from simple cells to complex multicellular ones, are capable of moving. Finding a way to change orientation and position in space is often the most important evolutionary mechanism for interacting with the environment: organisms could escape from unfavorable conditions or take nutrients, a bigger amount of light, etc. from surroundings. At the cellular level of life organization motility is often a determining factor of effective functionality for many types of cells. Bacteria locomote to gain organic and non-organic substances to support the life processes. Cells colonies organized movement depending on concentrations of different substances in the media is common for many bacterial species such as \textit{Salmonella enterica}. Chemotaxis \cite{Adler} - cell migration towards or away from the chemicals, chemoattractant, or chemorepellents,  dissolved in the cell's habitat, is being widely investigated today because of many potential applications of this process \cite{Karmakar}. Cells utilize receptors on their membrane to detect chemical gradients and employ shape changes to direct their locomotion. This effect is observed in many works, for instance, in \cite{Wong}.

For a long time, scientists distinguished just two types of cellular motility: flagellar motility and actin-dependent cell
migration \cite{Fritz-Laylin}, but recently the third type of cell motility in three-dimensional environments was described \cite{Yamada}. Flagellar motility is a type of movement in which a cell uses whip-like organelles that propel a cell through liquids by wave-like beating or rotational activity.  
Actin-dependent cell migration or crawling is another form of cellular activity that involves interaction with one or several surfaces. During the crawling, a cell produces protrusions on the membrane - lamellopodia, filopodia, pseudopodia , and blebs \cite{Fackler}, \cite{Paluch}, which can dynamically interact with the surface, clinging to it. The cell pulls itself up by the fixed area and, as a result, moves.

However, there are still a lot of challenges in the investigation and modeling of locomotion on 2D surfaces. Migration in real-world live 3D environments, such as tissues, complex protein structures, and fibrils, has been investigated even less \cite{Yamada}\cite{Petrie}. When describing such a movement, it is often necessary to take into account the heterogeneous structure of this medium and the interaction of the cell with it. Besides, new properties of cell motility appear when cells move in these environments. For example, when germ cells move in the process of embryo formation, the cell is capable of greatly deforming the surrounding space, that is, changing the structure of germ tissues, and forming pores for advancement. Cells often change the environment to enhance their ability to move. In \cite{Chi} authors mention that  \textit{Bacillus subtilis} speed of backward traveling on the same trajectory is larger than forward. That happens because a bacterium, while it moves forward, makes a tunnel in the liquid with liquid crystal properties, which facilitates swimming on the same path in the opposite direction.

In this paper, we consider the actin-dependent migration of a single cell. 
Crawling on the surface is a process that requires the involvement of a lot of molecular complexes. The main proteins \cite{Mogilner} engaged in the cell motility process are G-actin and myosin II. 
Actin is a globular protein that polymerizes on filaments. Myosin\cite{Shiroguchi} is a motor protein that could freely diffuse in the cytoplasm and interact with the actin sequences: it binds to the filaments and moves across them. The whole biochemical review of participating proteins can be found, for instance, in \cite{Li}.

The movement of the cell is complexly organized, and there is still no full understanding of many details of this process. However, it is well-known that there are three coordinated stages of cellular movement along the substrate \cite{Heidemann}: 
\begin{enumerate}
\item The extension of the cell membrane appears at the front of the cell, this process is called protrusion \cite{Clainche}. Recently, the protruded area of the cell has been considered to be driven only by the polymerization of the actin, but, nowadays, it is assumed that there is another protrusion called blebbing which occurs without polymerization \cite{Fackler}. Structures built during the actin-dependent protrusion are called lamellipodia, filopodia, or pseudopodia depending on the form of the protrusion \cite{Alblazi},\cite{Alberts}.
\item Subsequently, adhesion sites on the protruded area become activated. They are represented by the complex multi-protein assemblies known as focal adhesions (FAs), which mediate cell signaling, force transduction, and adhesion to the substratum \cite{Fraley},\cite{J. H.-C. Wang}, link the extracellular matrix via transmembrane matrix receptors to the actin cytoskeleton. 
During the crawling, focal adhesions assemble on the front protruded part of the cell and, thus, disassemble on the rear one. These adhesion complexes transmit traction forces that influence cell locomotion to the extracellular matrix. Cell motility is linked to the organization of the actin cytoskeleton \cite{Gardel}. It was demonstrated that its velocity is correlated linearly with the traction force values. 
According to this paper, the centripetal flow from the leading edge of a membrane is being formed. The flow called retrograde interacts with the FAs and drives the cell motility. To explain this phenomenon, the "molecular clutch" hypothesis was introduced in \cite{Case}.  
\item To move the center of the mass, a cell needs also to pull the back part forward. The process of pulling the rear part of the cell is called retraction. 
There exist several explanations of this process \cite{Cramer}. Moreover, the pulling of the membrane is carried out by a few mechanisms simultaneously. The sarcomere-like mechanism and transport-track mechanism depend on the interaction between actin and myosin: molecules of myosin bind both to the membrane and actin and pull the cell. Other mechanisms are only actin-dependent because they are based on the depolymerization of the filament or the recrosslinking of the shorted actin filament. 
Also, some studies show that retraction can be caused by the membrane tension \cite{Cramer}, \cite{Bershadsky}
However, the whole list of processes corresponding to the retraction is not full, and, probably, newer mechanisms will be described over time. 
\end{enumerate}

\section{{Theoretical background}}
The current work is based on the two-dimensional couple model described by the system of PDEs in the field of the dynamic domain $\Omega(t)$ with free boundary $\partial \Omega(t)$ for eukaryotic cells
on substrates \cite{Safsten}.
The key components of this model are Darcy's law, which describes the overloaded motion of the cytoskeleton of the active gel, and the Young-Laplace equation for pressure and velocity with an elastic nonlocal resistance force that depends on the area of a cell. The non-local term describing the pressure generated by the elastic force is
\begin{equation} \label{resistance force}
\begin{split}
p_{e}(|\Omega(t)|) = -k \left(\frac{|\Omega(t)|}{|\Omega_0|} - 1 \right),
 \end{split}
\end{equation}
where $k$ is the membrane stiffness. It is the generalization of the one-dimensional nonlocal spring condition introduced in
the works \cite{Putelat}, \cite{Recho}.
\\
The system consists of two PDEs and three boundary conditions. \\
The Darcy's Law for actomyosin:
\begin{equation} \label{Darcy}
\begin{split}
\Delta \varphi = \zeta \varphi - Q(m) \quad in \: \Omega(t).
 \end{split}
\end{equation}
The equation for the myosin advection-diffusion:
\begin{equation} \label{Diffusion}
\begin{split}
\frac{\partial m}{\partial t} + \nabla(m \nabla \varphi)- \Delta m = 0 \quad in \: \Omega(t).
 \end{split}
\end{equation}
The Young-Laplace equation:
\begin{equation} \label{YoungLaplace}
\begin{split}
\zeta \varphi = -\gamma\kappa + p + p_{e} (|\Omega(t)|) \quad on \: \partial \Omega(t).
 \end{split}
\end{equation}
The continuity of velocities on the boundary: 
\begin{equation} \label{TangentCont}
\begin{split}
V_{\nu} = \partial_{\nu} \varphi \quad on \: \partial \Omega(t).
 \end{split}
\end{equation}
The non-flow condition: 
\begin{equation} \label{Non-flow}
\begin{split}
\partial_{\nu} m = 0 \quad on \: \partial \Omega(t).
 \end{split}
\end{equation}
There $\varphi$ is a potential of the actomyosin gel, $m$ is a myosin density, $\zeta$ and $\gamma$ are coefficients of the adhesion and the surface tension, respectively, $p$ is a constant homeostatic pressure, $\kappa$ is a curvature. $\nu$ is a normal vector to the boundary and $V_{\nu}$ is a border velocity. Compared to the initial statement of the problem the myosin density in (\ref{Darcy}) has been replaced by the normalizing function $Q(m)=\frac{1-e^{-m}}{1+e^{-m}}$.

The explicit derivation of the equation system of the model can be found in \cite{Safsten}-\cite{Recho}. 

\section{Numerical solution}

\subsection{Polar coordinates}
\begin{figure}[h]
\centering
\includegraphics[width=2.5in]{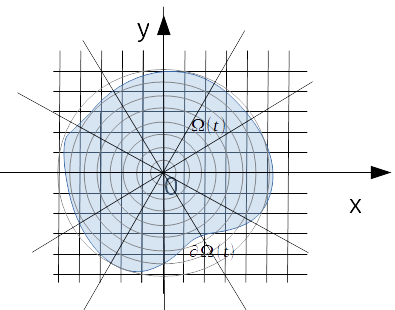}
\caption{Polar coordinates change}
\label{Fig:1}
\end{figure}

We first transform the system to the polar coordinates (Fig. \ref{Fig:1}).

For the equations (\ref{Darcy}) and (\ref{Diffusion}) that are defined in the domain,   $ \Omega(t)$ which in polar coordinates could be defined as $0<r \leq \rho(\theta,t)$: 
\begin{equation} \label{DarcyPolar}
\begin{split}
{\varphi _{rr}} + \frac{1}{r}{\varphi _r} + \frac{1}{{{r^2}}}{\varphi _{\theta \theta }} = \zeta \varphi  - Q(m),
 \end{split}
\end{equation}

\begin{equation} \label{DiffusionPolar}
\begin{split}
{m_t} = {m_{rr}} + \frac{1}{r}{m_r} + \frac{1}{{{r^2}}}{m_{\theta \theta }} - {m_r}{\varphi _r} - \frac{1}{{{r^2}}}{m_\theta }{\varphi _\theta } \\ - m\left( {\zeta \varphi  - Q(m)} \right).
 \end{split}
\end{equation}

The boundary conditions we assume here are  (\ref{YoungLaplace})-(\ref{Non-flow}) on $ \partial \Omega(t)$ which could be denoted now as $r = \rho(\theta, t)$ transform to the following:

\begin{equation} \label{YoungLaplacePolar}
\begin{split}
\zeta \varphi  = p - k\left[ { \frac{|\Omega(t)|}{\pi R_0^2} - 1} \right] - \gamma \kappa,
 \end{split}
\end{equation}

where $\kappa = \frac{{\rho ^2} - \rho {\rho _{\theta \theta }} + 2{\rho ^2 _\theta }}{{\left( {{\rho ^2} + { \rho _\theta^2}} \right)}^{3/2}}$ and $|\Omega(t)| = \frac{1}{2} \int_{ - \pi }^\pi  {{\rho ^2}d\theta } $,

\begin{equation} \label{TangentContPolar}
\begin{split}
{\rho _t} = {\varphi _r} - \frac{{{\rho _\theta }}}{{{\rho ^2}}}{\varphi _\theta },   
 \end{split}
\end{equation}
\begin{equation} \label{Non-flowPolar}
\begin{split}
{\rho ^2}{m_r} - {\rho _\theta }{m_\theta } = 0.   
 \end{split}
\end{equation}

\begin{figure}[h]
\centering
\includegraphics[width=2.5in]{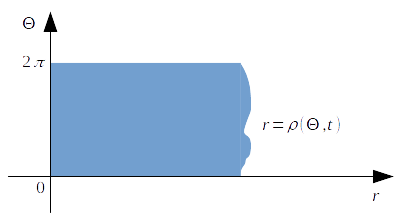}
\caption{Area of the cell in the system $(r, \theta$)}
\label{Fig:2}
\end{figure}

In the $(r, \theta)$ coordinates a domain of the cell can be represented by the rectangle with one curvilinear side which is a graph of the function $r = \rho(\theta, t)$ (Fig. \ref{Fig:2}), boundary defining function to be identified. The case of the system in polar coordinates is the common one due to the singularity of coefficients in the center of coordinates $r = 0$. It should be dealt with in the finite difference scheme which is described in section D.

\subsection{Front-fixing transformation of the coordinate system}

Converting the system (\ref{DarcyPolar})-(\ref{Non-flowPolar}) to the coordinates $(\hat r, \theta)$, where $\hat r = \frac{r}{\rho(\theta, t)}$, similar to \cite{Vabishevich} and \cite{Golosnoy} we obtain the equations  that are transformed to the following: 

\begin{equation}
\begin{split}
{\hat r}^{2}{\varphi _ {\hat r \hat r}}{\sin {\omega} } + \left( {\varphi _{\theta \theta }} - {\hat r ^2} {\rho ^2}(\zeta \varphi  - Q(m))\right){\sin^{3} {\omega} } - \\ - {\hat r}{\varphi _{\hat r \theta }}{\sin {\omega} }{\sin 2{\omega} }  +{\hat r}\rho \kappa {\varphi _{\hat r}} = 0,
 \end{split}
 \label{DarcyCurviPolar}
\end{equation}

\begin{equation}
\begin{split}
\rho ^2 m_t = \frac{1}{\sin^2{\omega}} \left({m_{\hat{r} \hat{r}} - m_{\hat r} \varphi_{\hat r}}\right) +\frac{1}{\hat r ^2} \left(m_{\theta \theta} - m_{\theta} \varphi_ \theta \right) + \\ + \frac{\cot \omega}{\hat r} \left(m_{\theta} \varphi _{\hat r} - m_{\hat r} \varphi _{\theta} - 2 m_{\hat r \theta} \right) + \frac{\kappa \rho}{\hat{r} \sin^3{\omega}}m_{\hat{r}} -\\
- \rho ^2 m \left( \zeta \varphi - Q(m)\right),
 \end{split}
  \label{DiffusionCurviPolar}
\end{equation}

where $\omega = \omega(\theta)$ is the angle between the tangent and radius vector at the boundary point $(\rho (\theta , t), \theta)$ and $\sin{\omega}=\frac{\rho}{\sqrt{\rho^2 + \rho_\theta^2}}$. The boundary conditions:

\begin{equation}
\begin{split}
\zeta \varphi  = p - k\left[ { \frac{|\Omega(t)|}{{\pi R_0^2}}  - 1} \right] - \gamma \kappa, 
 \end{split}
 \label{YoungLaplaceCurviPolar}
\end{equation}
\begin{equation}
\begin{split}
{\rho _t} = \frac{1}{\rho }{\varphi _{\hat{ r} }} - \frac{{{\rho _\theta }}}{{{\rho ^2}}}\left( {{\varphi _\theta } - \frac{{\hat{r} {\rho _\theta }}}{\rho }{\varphi _{\hat{r} }}} \right) ,
 \end{split}
 \label{TangentContCurviPolar}
\end{equation}
\begin{equation}
\begin{split}
\rho {m_{\hat{r} }} - {\rho _\theta }\left( {{m_\theta } - \frac{{\hat{ r} {\rho _\theta }}}{\rho }{m_{\hat{ r} }}} \right)  = 0.
 \end{split}
 \label{Non-FlowCurviPolar} 
\end{equation}

The domain of the problem in $(\hat r, \theta)$ coordinates is the rectangle $[0;1]\times [0;2\pi]$. The boundary conditions (\ref{YoungLaplaceCurviPolar})-(\ref{Non-FlowCurviPolar}) are valid for the side $\hat r = 1$ of the rectangle. For the other sides of the domain, consider the periodicity and zero node conditions:

\begin{equation} \label{Periodicity}
m (0, \hat r) = m (2\pi, \hat r),{\ } \varphi (0, \hat r) = \varphi (2\pi, \hat r),
\end{equation}
\begin{equation} \label{ZeroNode}
m (0,0) = m (\theta ,0),{\ } \varphi (0,0) = \varphi (\theta ,0).
\end{equation}

Further, we are going to split the problem into the smaller ones: the boundary value problem for potential $\varphi(\hat r, \theta, t)$; the myosin $m(\hat r, \theta, t)$ diffusion and the border function's $\rho(\theta, t)$ evolution.

\subsection{The finite difference scheme of the boundary value problem for the potential $\varphi$ approximation}

Suppose the moment of time $t$ is fixed. Equation (\ref{DarcyCurviPolar}) together with boundary conditions  (\ref{YoungLaplaceCurviPolar}), periodicity (\ref{Periodicity}) and zero node conditions (\ref{ZeroNode}) form the mixed boundary value problem for the potentials of the velocity's field. We build a finite difference scheme for this problem on the uniform grid in $(\hat r, \theta)$ coordinates: $\left\{ {{r_i} = hi:i \in \mathbb{Z},0 \le i \le N,h = 1/N} \right\}$, $\left\{ {{\theta _j} = \eta j:j \in \mathbb{Z},0 \le j \le 4M,\eta  = \pi /2M} \right\}$, where $\varphi_{i,j} = \hat{\varphi}({r_i},{\theta _j}, t)$, $m_{i,j} = \hat{m}({r_i},{\theta _j}, t)$, $\rho_j = \hat \rho(\theta_j,t)$, $\rho^{'}_j = \hat \rho_{\theta}(\theta_j,t)$, $\rho^{''}_j = \hat \rho_{\theta \theta}(\theta_j,t)$, $ \hat{\varphi}$ - an approximate solution of the system (\ref{DarcyCurviPolar}), (\ref{YoungLaplaceCurviPolar}), (\ref{Periodicity}), (\ref{ZeroNode}), $\hat{m}$ is the approximation of the current myosin distribution; $\hat \rho$, $\hat \rho _{\theta}$,  $\hat \rho _{\theta \theta}$ are approximations of the current state of the boundary function and its derivatives. Then, the second-order finite difference scheme is used: 
\begin{equation}, 
\begin{split} \label{MainFiniteDifference}
B_{ij}(\varphi_{i+1,j+1} - \varphi_{i+1, j - 1} - \varphi_{i-1,j+1} + \varphi_{i-1, j-1})+ \\ + (2A_{ij} + 2C_{j} + {r^2_i}{\rho^2_j}\zeta {\sin^3 {\omega_j}}) \varphi_{i,j} - (A_{ij}+D_{ij})\varphi_{i+1,j}- \\ - (A_{ij}-D_{ij}) \varphi_{i-1,j} - C_j \left(\varphi_{i,j+1} + \varphi_{i,j-1} \right) = \\ = {r^2 _i} \rho^2_j Q(m_{i,j}) {\sin^3 \omega_j},
 \end{split}
\end{equation}

where  ${\sin {\omega_j}} = \frac{\rho_j}{\sqrt{{\rho^2_j} + \rho ^{'2}_j}}$, ${A_{ij}} = \frac{{r^2_i} \sin {\omega_j}}{h^2}$, ${B_{ij}} = \frac{2{r_i} \sin^2 {\omega_j} \cos {\omega_j}}{h \eta}$, ${C_{j}} = \frac{\sin^3 {\omega_j} \cos {\omega_j}}{\eta^2}$, ${D_{ij}} = \frac{\kappa_j \rho_j {r_i} }{2h}$,  $\kappa_j = \frac{{\rho^2_j} - \rho_j {\rho^{''} _{j }} + 2{\rho^{'2}_j }}{{\left( {{\rho^2_j} + { \rho^{'2}_j}} \right)}^{3/2}}$.

\begin{figure}[h]
\centering
\includegraphics[width=3.5in]{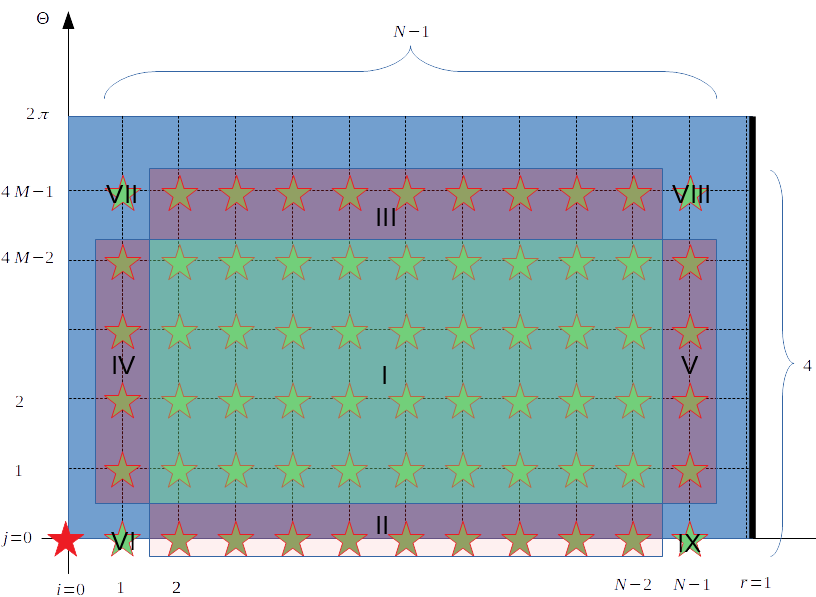}
\caption{Classification of the grid nodes}
\label{NodeClasses}
\end{figure}

The grid nodes are divided into 10 classes (Fig. \ref{NodeClasses}). Class I is the main one, for nodes of which the main difference scheme is used. For Classes II and III nodes the main difference scheme is changed according to the periodicity condition (\ref{Periodicity}) which are $\varphi_{i,0} = \varphi_{i,4M}$ and $m_{i,0} = m_{i,4M}$. Class IV are nodes for which the main finite difference scheme (\ref{MainFiniteDifference}) is added with the zero node condition (\ref{ZeroNode}) which is $\varphi_{0,j}=\varphi_{0,0}$. Class V contains nodes for which the main difference scheme is used in combination with the boundary condition (\ref{YoungLaplaceCurviPolar}) which is $\zeta \varphi_{N,j} = p - k \left[ \frac{|\Omega(t)|}{{{\pi R_0^2}}  - 1} \right] - \gamma \kappa_j $. The boundary condition (\ref{YoungLaplaceCurviPolar}) defines the right-hand side of the final finite difference linear system. Classes VI and VII contain one node each for which the periodicity condition (\ref{Periodicity}) and the zero node condition (\ref{ZeroNode}) are used to complete the main finite difference scheme (\ref{MainFiniteDifference}). Classes VIII and IX contain one node each for which the periodicity conditions (\ref{Periodicity}) and the boundary condition (\ref{YoungLaplaceCurviPolar}) are used to complete the main finite difference scheme (\ref{MainFiniteDifference}). 
As a result, we have the grid with $(N-1)\times(4M-1)+1$ independent nodes which are marked by the five-pointed stars (Fig. \ref{NodeClasses}). The finite difference scheme described above provides the system of $(N-1)\times(4M-1)$ linear equations. To complete the system, we have to build an approximation of the Laplace operator on $\varphi$ in the center of coordinates. It has not been built yet due to singularities in coefficients of equations in polar coordinates.

\subsection{Finite difference scheme for zero node $\hat r = 0$}
\begin{figure}[h]
\centering
\includegraphics[width=2.5in]{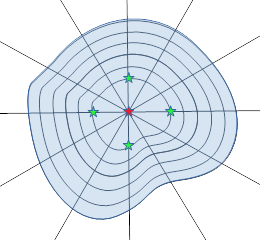}
\caption{Uneven grid for the Laplacian approximation}
\label{UnevenGrid}
\end{figure}

Coefficients of the equation (\ref{DarcyCurviPolar}) have a singularity at the point $\hat r = 0$ as well as those of the equation (\ref{DarcyPolar}) at the point $r = 0$. Therefore, to provide the finite difference scheme for the zero node we return to the initial form of the equation (\ref{Darcy}) remaining in the grid determined above. In coordinates $(x, y)$ the grid is uneven (Fig. \ref{UnevenGrid}). Indeed, Cartesian coordinates for nodes on the x-axis are $(-h\rho(\pi, t); 0)$ and $(h\rho(0, t); 0)$ and for nodes on the y-axis are $(0; h\rho(\pi/2, t),)$ and $(0; -h\rho(3\pi/2, t))$. 

The finite difference scheme for the approximation of the second derivative on an uneven grid is described in \cite{Samarskiy}. According to that,

\begin{equation}
\begin{split} \label{UnevenApproximationXX}
\frac{1}{h_0}\left( \frac {\varphi_{1,0}-\varphi_{0,0}}{h\rho_0} - \frac {\varphi_{0,0}-\varphi_{1,2M}}{h\rho_{2M}}  \right) = \\
=\varphi_{xx}(0;0)+\frac{h(\rho_{0}-\rho_{2M})}{3}\varphi_{xxx}(0;0) + O(h^2),
 \end{split}
\end{equation}

\begin{equation}
\begin{split} \label{UnevenApproximationYY}
\frac{1}{h_{M}}\left( \frac {\varphi_{1,M}-\varphi_{0,0}}{h\rho_M} - \frac {\varphi_{0,0}-\varphi_{1,3M}}{h\rho_{3M}}  \right) = \\
=\varphi_{yy}(0;0)+\frac{h(\rho_{0}-\rho_{2M})}{3}\varphi_{yyy}(0;0) + O(h^2),
 \end{split}
\end{equation}
where $h_0 = 0.5h(\rho_0+\rho_{2M})$ and $h_{M} = 0.5h(\rho_M+\rho_{3M})$. The sum of (\ref{UnevenApproximationXX}) and (\ref{UnevenApproximationYY}) gives us an approximation of Laplacian at the coordinate center at the moment t. Using this approximation and the initial equation (\ref{Darcy}) for the zero node, we obtain the last linear equation for the finite difference scheme. The drawback of the approximation is the first order of accuracy. To improve the order of approximation, we build several approximations of the Laplacian at the zero node. To do that, denote $h_j = 0.5h(\rho_j+\rho_{j + 2M})$ and $B_j = \frac {h(\rho_j - \rho_{j + 2M})}{3}$. The first order approximation of the second directional derivative along the vector ${\vec{v}_j} = (\cos \theta_j, \sin \theta_j)$ on the uneven grid is denoted as follows:
\begin{equation}
\begin{split} \label{UnevenApproximationV}
D^2_j \varphi = \frac{1}{h_j}\left( \frac {\varphi_{1,j}-\varphi_{0,0}}{h\rho_j} - \frac {\varphi_{0,0}-\varphi_{1,j + 2M}}{h\rho_{j + 2M}}  \right)
 \end{split}
\end{equation}

\begin{figure}[h]
\centering
\includegraphics[width=2.5in]{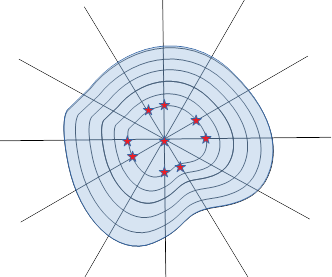}
\caption{Two cross-patterns rotated relative to each other with center at the zero node}
\label{RotatedCrossPatterns}
\end{figure}

Then, using the invariance of the Laplace operator with respect to the rotation of the coordinate system, we obtain a family of its approximations:
\begin{equation}
\begin{split} \label{UnevenApproximationFam}
D^2_j \varphi + D^2_{j+2M} \varphi = \\ 
= \Delta \varphi + B_j C_j (\varphi) + B_{j+2M} C_{j+2M} (\varphi) + O(h^2)
 \end{split}
\end{equation}
where $C_j (\varphi)=\varphi_{xxx} \cos ^3 \theta_j+3 \varphi_{xxy} \cos ^2 \theta_j \sin \theta_j+3 \varphi_{xyy} \cos \theta_j \sin ^2 \theta_j+\varphi_{yyy} \sin ^3 \theta_j$, all values of third derivatives of function $\varphi$ are taken in the coordinate center $(0, 0)$. All approximations from the family are approximations of the Laplace operator on $\varphi$ in the coordinate center on the cross-patterns rotated relative to each other (Fig. \ref{RotatedCrossPatterns}). 

The approximations are of the first order, however, coefficients of the first-order term in the residual of an approximation are homogeneous polynomials of four variables $(\varphi_{xxx}, \varphi_{xxy}, \varphi_{xyy}, \varphi_{yyy})$ with known coefficients given in the definitions of $B_j$ and $C_j$. Therefore, using their non-trivial linear combination we eliminate the first-order term. To do this, we choose $M = 5m$ and build five approximations from the family rotated by $\pi / 10$ relative to each other, calculate five vectors of corresponding coefficients of polynomials determining the first order terms of residuals for each approximation, find the non-trivial linear combination equal to zero for the set of the five vectors using the least square method, then use the coefficients to form second-order approximation as a linear combination of the corresponding approximations from the family.

In more detail, denote vectors 
\begin{equation}
\begin{split} \label{PolinomCoef}
a_l = \begin{pmatrix}
B_l \cos ^3 {\frac {l \pi} {10}} + B_{l+2M} \cos ^3 { \frac {(10 + l) \pi} {10}}\\
B_l \cos ^2 {\frac {l \pi} {10}} \sin {\frac {l \pi} {10}} + B_{l+2M} \cos ^2  \frac {(10 + l) \pi} {10}\sin { \frac {(10 + l) \pi} {10} }\\
B_l \cos {\frac {l \pi} {10}} \sin ^2 {\frac {l \pi} {10}} + B_{l+2M} \cos  \frac {(10 + l) \pi} {10}\sin ^2 { \frac {(10 + l) \pi} {10} }\\
B_l \sin ^3 {\frac {l \pi} {10}} + B_{l+2M} \sin ^3 { \frac {(10 + l) \pi} {10}}
\end{pmatrix}
 \end{split}
\end{equation}
and coefficients $\alpha_i$ for $i \in \begin{Bmatrix} 0, & 1, & 2, & 3, & 4 \end{Bmatrix}$ which present a non-trivial solution of the linear system

\begin{equation}
\begin{split} \label{SecondOrderCondition}
 \sum_{i=0}^{4} \alpha_i a_{i*m}  = 0
 \end{split}
\end{equation}

 The rank of the system (\ref{SecondOrderCondition}) is no more than 4 while there are 5 independent variables $\alpha_i$. Therefore, there exists solution of the system (\ref{SecondOrderCondition}) extended with the equation 

\begin{equation}
\begin{split} \label{SecondOrderConditionEx}
 \sum_{i=0}^{4} \alpha_i = 1
 \end{split}
\end{equation}
 
 Then, the second-order approximation for $\Delta \varphi (0,0)$ can be obtained as follows

\begin{equation}
\begin{split} \label{ZeroNodeLaplas}
\sum_{i=0}^{4} \alpha_i (D^2_{im} \varphi + D^2_{im+2M} \varphi) = \\ 
= \Delta \varphi + O(h^2)
 \end{split}
\end{equation}  

 Here the condition (\ref{SecondOrderCondition}) guaranties elimination of the first order residual (see (\ref{UnevenApproximationFam})) and the condition (\ref{SecondOrderConditionEx}) guarantees no additional multiplier for the $\Delta \varphi$ term in the right-hand side of (\ref{ZeroNodeLaplas}). 
 
 Finally, based on the known myosin distribution and the form of the boundary or their approximations at the current time, we obtain the approximation of potential $\varphi$ of the actomyosin complex velocity field. The numerical solution of the mixed boundary problem described by the system (\ref{DarcyCurviPolar}), (\ref{YoungLaplaceCurviPolar}), (\ref{Periodicity}), (\ref{ZeroNode}) is now reduced to the solution of the linear system of $(N-1)\times(4M-1)+1$ order.

\subsection{Myosin diffusion, boundary evolution, and the software structure}
The equation describing myosin diffusion (\ref{DiffusionCurviPolar}) is resolved with respect to the first time derivative. Consequently, by incorporating the boundary condition (\ref{Non-FlowCurviPolar}) and considering the current state of the system, we can calculate myosin diffusion in the next time step using either the explicit Euler scheme or the second-order implicit Runge-Kutta method \cite{Runge-Kutta}. The same is done for the $\rho(\theta, t)$ function describing the evolution of the boundary (\ref{TangentContCurviPolar}).
The issue is that the coordinate system $(\hat r, \theta)$ changes with the $\rho(\theta, t)$ function. In addition, the system coefficients become unstable the closer the coordinate system center is to the boundary. Therefore, the coordinate system, its center, and the grid should be adapted to the domain of the system. The coordinate center moves to the center of mass of the evolved boundary, consequently, on each cycle. The new approximation grid is built and the functions $\varphi (\hat r, \theta)$ and $m (\hat r, \theta)$ are interpolated for the new nodes of the evolved grid using a cubic spline interpolation.

\section{Software tests and numerical results}
\begin{figure}[h]
\centering
\includegraphics[width=3in]{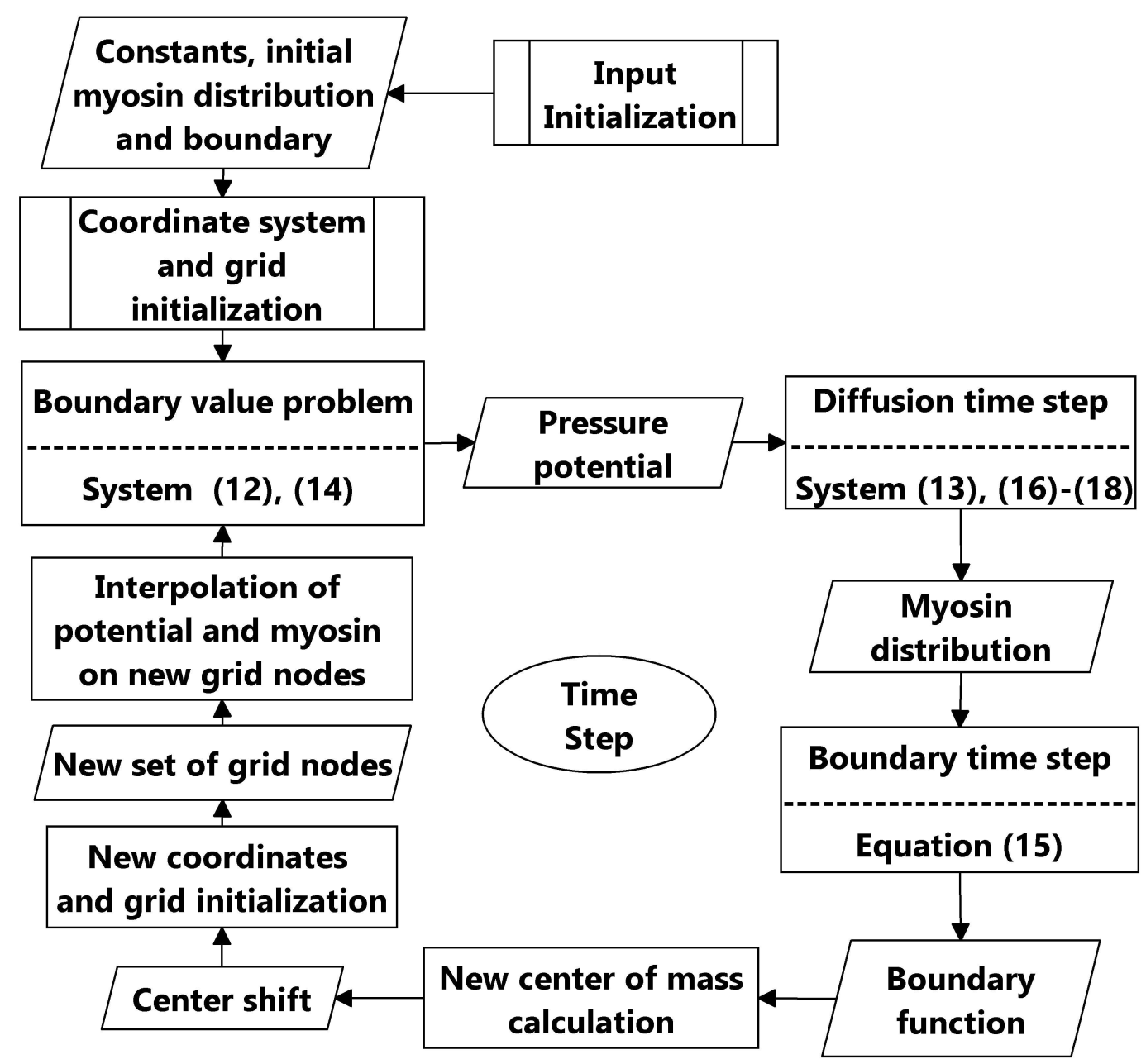}
\caption{Block scheme of the calculation workflow}
\label{BlockScheme}
\end{figure}

To implement the described algorithm (Fig.\ref{BlockScheme}), a Python software package was developed. 

The software module for the numerical solution of the boundary value problem (\ref{Darcy}, \ref{YoungLaplace}) described in Section III.B-D was tested on the family of static solutions $\varphi=C_1 e^{\sqrt{\zeta}x}+C_2 e^{\sqrt{\zeta}y}, m=0$ without the myosin diffusion and border dynamic for an arbitrary boundary.

Static solution 
\begin{equation} \label{StaticPhi}
\begin{split}
\varphi=\frac{p}{\zeta}+\frac{k}{\zeta} \left( \frac{R^2}{R^2_0} - 1 \right) - \frac{\gamma}{\zeta R},
 \end{split}
\end{equation}

\begin{equation} \label{StaticM}
\begin{split}
 m={p}-{k} \left( \frac{R^2}{R^2_0} - 1 \right) - \frac{\gamma}{R}
 \end{split}
\end{equation}
on the round area $\rho(\theta, t)=R$ has also been tested.

A family of dynamic radial-symmetric solutions $\varphi(r, t)$ was also considered. We assumed that $m = 0$ and that the initial border shape is a circle $\rho(\theta, 0) = \rho_0$. Then, from the (\ref{Darcy}), (\ref{YoungLaplace}), (\ref{Non-flow}) and taking into account the initial condition, we get the following system: 

\begin{equation} \label{bessel equation}
\begin{split}
r \varphi_r + r^2 \varphi_{r r} =\zeta r^2 \varphi  \quad |    0<r<\rho(t),
 \end{split}
\end{equation}

\begin{equation} \label{border-bessel}
\begin{split}
\zeta \varphi = - \frac{\gamma}{\rho} + p - k\left(\left(\frac{\rho}{\rho_0}\right)^2 - 1 \right)  \quad  |    r = \rho(t),
 \end{split}
\end{equation}

\begin{equation} \label{velocities bessel}
\begin{split}
\varphi_r = \rho_t \quad |    r = \rho(t),
 \end{split}
\end{equation}

\begin{equation} \label{initial condition}
\begin{split}
\rho(0) = \rho_0.
 \end{split}
\end{equation}

The solution of the (1) is $\varphi(r,t) = C(t)I_0(\sqrt{\zeta} r)$, where $I_0$ is the modified Bessel function of the first kind, $C(t) = \frac{- \gamma \rho_0^2 + p \rho_0^2 \rho(t)  - k (\rho_0^2 - \rho(t)^2) \rho(t) +}{I_0(\sqrt{\zeta}\rho(t)) \zeta \rho(t) \rho_0^2}$. $\rho(t)$ is a solution of the Cauchy problem (\ref{C1})-(\ref{C2}). 

\begin{equation} \label{C1}
\begin{split}
\rho_t = f(t),
 \end{split}
\end{equation}

\begin{equation} \label{C2}
\begin{split}
\rho(0) = \rho_0,
 \end{split}
\end{equation}

where 
\begin{equation} \label{RightPart}
\begin{split}
f(t) = C(t) \frac{d I_0(\sqrt{\zeta} r)}{d r} \bigg|_{\rho(t)}
 \end{split}
\end{equation}

Thus, functions $\rho(t)$, $C(t)$, and $\varphi (t)$ can be found by Euler or Runge-Kutta methods.

\begin{figure}[h]
\centering
\includegraphics[width=2.55in]{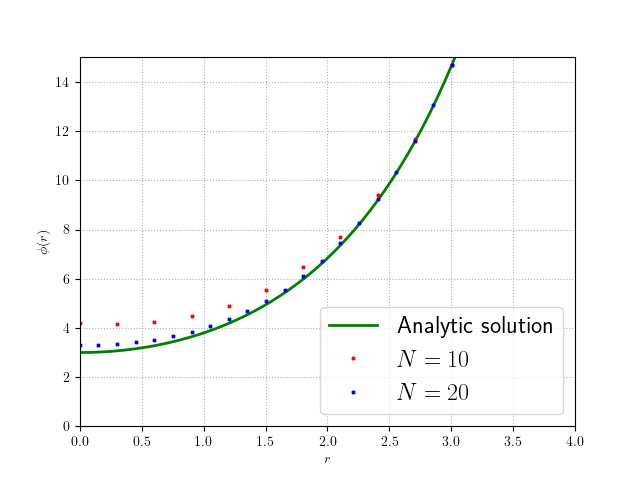}
\caption{The graph of function $\varphi(r)$ for $\rho_0 = 2$, $t=0.25$, $\zeta = 0.2$, $\gamma = 0.05$, $R_0 = 2$, $k = 0.25$ (green curve is the analytic solution, red dots represent the numerical solution for $N=10$, blue dots is the numerical solution for $N=20$) }
\label{BesselAssimptotic}
\end{figure}

Numerical tests of the whole software package with the circular initial cell form and $m=0$ have been conducted with different grid steps to test the convergence of the numerical solution to the analytic one and asymptotic of the convergence. The numerical results show convergence to the static solution described above with the circle boundary of radius $R$ which is a solution of the equation (\ref{StaticM}) for $m=0$. Depending on initial radius $\rho_0$, the cell area shrinks or stretches to the circle of radius $R$ without oscillations while function $\varphi(r,t)$ converges to zero in every point of the area. The time of convergence is $T = 2.5$. The residual of the numerical solution shows $\sim \frac {1}{N^2}$ asymptotic (Fig. \ref{BesselAssimptotic}). The calculation time for $N=20$, $M = 10$ takes $~0.5$ second for every time layer on the laptop with i5-7200U CPU.

The radial-symmetric initial conditions and corresponding numerical solution show no dependence from $\theta$ and, therefore, can not be used to evaluate the residual dependence on $M$. However, we used the same family of initial conditions and analytical solutions in the coordinate system with shifted center while center shift modules were turned off (see Fig.\ref{BlockScheme}) during the calculations. The residuals showed $ O(\frac {1}{N^2}) + O(\frac {1}{MN}) + O(\frac {1}{M^2})$ asymptotic.

\begin{figure}[h]
    \centering

    \begin{subfigure}{0.8\linewidth}
        \centering
        \includegraphics[width=0.71\linewidth]{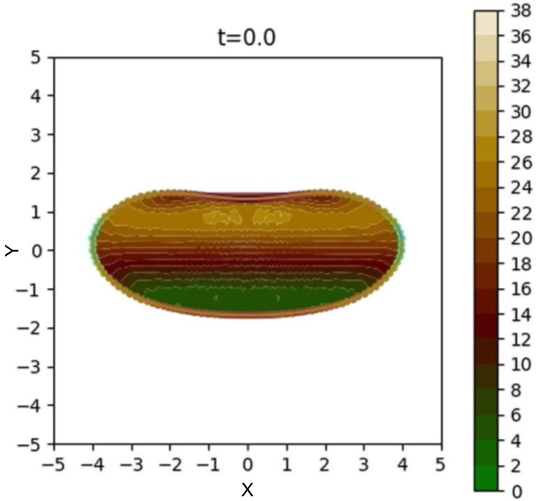}
        \caption{The initial shape of the keratocyte at the start of the dynamic process.}
        \label{fig:a}
    \end{subfigure}

    \begin{subfigure}{0.8\linewidth}
        \centering
        \includegraphics[width=0.71\linewidth]{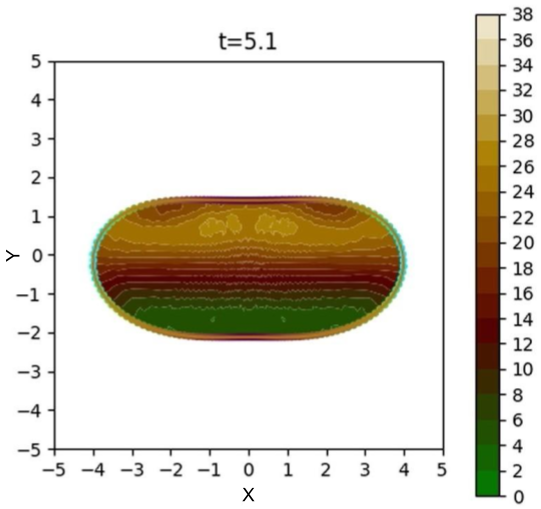}
        \caption{The shape change phase. The initial cell shape and myosin distribution are close to the stable solution of traveling wave type and tend to take the configuration for the migration phase.}
        \label{fig:b}
    \end{subfigure}

    \begin{subfigure}{0.8\linewidth}
        \centering
        \includegraphics[width=0.71\linewidth]{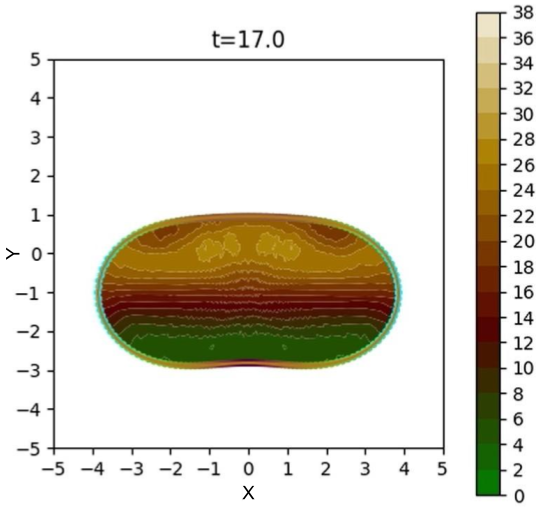}
        \caption{The migration phase. The cell moves at a constant speed in a straight line with almost no changes in shape and actomyosin complex distribution.}
        \label{fig:c}
    \end{subfigure}

    \caption{The myosin distribution and cell shape dynamics of a keratocyte.}
    \label{fig:KeratoDynamic}
\end{figure}

Authors of \cite{Safsten} hypothesized the existence of stable traveling wave solutions for the (\ref{Darcy})-(\ref{Non-flow}) free-boundary problem, which could be a confirmation of the applicability of the presented model for cell motility mechanism. We digitized the actomyosin distribution and shape of moving keratocyte cells observed in \cite{Barnhart} and used it as an initial condition of the described dynamic system (Fig. \ref{fig:a}). The observed behavior of the cell in the numerical experiment showed two phases of evolution: shape change(\ref{fig:a}, \ref{fig:c}) and uniform rectilinear motion with almost no change in shape and actomyosin complex distribution(\ref{fig:b}), which can be interpreted as stable traveling wave solution.  

\section{Conclusions}
The phenomenon of chemo-taxis of eukaryotes is closely related to actin-dependent cell motility. The size of a single cell does not allow it to determine the local gradients of chemoattractant concentration at each moment of time. Therefore, every cell enters the cultivation phase in the presence of an unevenly distributed chemoattractant. In the process of cultivation, colonies containing several thousand cells are formed and acquire a characteristic polarized shape. The active cell migration phase begins when the cell shape and actomyosin distribution in it take a particular configuration. Typical moving cell shapes are elongated in the direction perpendicular to the direction of movement and have convex(leading) and concave (driven) edges or vice versa. The average migration rate is several cell sizes per hour \cite{Wong}. It is shown in this paper that the processes of cell shape change and its transition to the migration phase, observed in real experiments can be described using the cell motility model proposed earlier in \cite{Safsten}. The stable dynamic solutions of a traveling wave type exist.

The developed numerical scheme and corresponding software showed stable results which were tested on analytic solutions. The software should be further developed and supposed to be used as an instrument of the investigation and further development of the model of cell motility. The next step is supposed to be an introduction of the surrounding medium as a viscoelastic liquid with chemoattractant distribution, exchange of non-flow boundary condition for the myosin concentration to the diffusion boundary condition and take into account the surrounding liquid flow and myosin diffusion.

\scenario*{Contribution of individual authors to the creation of a scientific article (ghostwriting policy)}
Vitaly Chernik designed the numerical solution scheme and overall software architecture.
Pavel Buklemishev carried out the literature review, structured the scientific background of the model, developed software, and carried out the simulations and code optimization.\\


\begin{thebibliography}{1}

\bibitem{Adler} J. Adler. (1966, August). Chemotaxis in bacteria. Science. 153(3737): pp. 708-716. doi: 10.1126/science.153.3737.708. 
%??????? ? ???, ????????, ??? ?????, ? ??????? ?? ?????
%edited
\bibitem{Karmakar}
R. Karmakar. (2021, May). State of the art of bacterial chemotaxis. J Basic Microbiol. 61(5): pp. 366-379. doi: 10.1002/jobm.202000661.
%edited
\bibitem{Wong}C.-W. Wong,  C. F. LeGrand, B. F.  Kinnear,  et al. (2019, December). In Vitro Expansion of Keratinocytes on Human Dermal Fibroblast-Derived Matrix Retains Their Stem-Like Characteristics. Sci Rep 9, 18561.doi:10.1038/s41598-019-54793-9.
%?????? 6 ??????? et al
%????????? ?????? ??????
\bibitem{Fritz-Laylin}
L. K. Fritz-Laylin(2020 May). The evolution of animal cell motility. Current Biology. 2020 May 18;30(10):R477-82. doi:10.1016/j.cub.2020.03.026.
 \bibitem{Yamada}
   K. M. Yamada, M. Sixt. (2019, October). Mechanisms of 3D cell migration. Nat Rev Mol Cell Biol 20, 738-752 (2019).
\bibitem{Fackler}
O. T. Fackler, R. Grosse.(2008, June). Cell motility through plasma membrane blebbing. J Cell Biol. 2008 Jun 16;181(6):879-84. doi: 10.1083/jcb.200802081. 
\bibitem{Paluch}
E. K. Paluch, E. Raz. (2013, June). The role and regulation of blebs in cell migration. Curr Opin Cell Biol. 2013 Oct;25(5):582-590. doi:10.1016/j.ceb.2013.05.005. 
\bibitem{Petrie}
R. J. Petrie, K. M. Yamada. (2016, April). Multiple mechanisms of 3D migration: the origins of plasticity. Curr Opin Cell Biol. 2016 Oct;42:7-12. doi: 10.1016/j.ceb.2016.03.025.
\bibitem{Chi}
H. Chi, A. Gavrikov, L. Berlyand, I. S. Aranson(2022, November). Interaction of microswimmers in viscoelastic liquid crystals. Commun Phys 5, 274 (2022). https://doi.org/10.1038/s42005-022-01056-1.
\bibitem{Mogilner}
A. Mogilner.(2008, May). Mathematics of cell motility: have we got its number? J. Math. Biol. 58, 105-134 (2009). https://doi.org/10.1007/s00285-008-0182-2
\bibitem{Shiroguchi}
K. Shiroguchi, K. Jr. Kinosita.(2007 May). Myosin V walks by lever action and Brownian motion. Science.  25;316(5828):1208-1212. doi: 10.1126/science.1140468. 
\bibitem{Li}
S. Li, J.-L. Guan, S. Chien. (2005, August). Biochemistry and biomechanics of cell motility. Annu Rev Biomed Eng. 2005;7:105-50. doi: 10.1146/annurev.bioeng.7.060804.100340. 
\bibitem{Heidemann}
S. R. Heidemann, R. E. Buxbaum. (1998, April). Cell crawling: first the motor, now the transmission. J Cell Biol. 1998 Apr 6;141(1):1-4. doi: 10.1083/jcb.141.1.1.
\bibitem{Clainche}
C. Le Clainche, M.-F. Carlier(2008, April). Regulation of actin assembly associated with protrusion and adhesion in cell migration. Physiol Rev. 2008 Apr;88(2):489-513. doi: 10.1152/physrev.00021.2007.

\bibitem{Alblazi}
K. M. Alblazi, C. H. Siar. (2015, June). Cellular protrusions--lamellipodia, filopodia, invadopodia and podosomes--and their roles in progression of orofacial tumours: current understanding. Asian Pac J Cancer Prev. 2015;16(6):2187-91. doi: 10.7314/apjcp.2015.16.6.2187. 
\bibitem{Alberts}
B. Alberts, A. Johnson, J. Lewis, M. Raff, K. Roberts, P. Walter. Molecular Biology of the Cell. 4th edition. New York: Garland Science; 2002. The Cytoskeleton and Cell Behavior. Available from: https://www.ncbi.nlm.nih.gov/books/NBK26930/
\bibitem{Fraley}
S. I. Fraley, Y. Feng, R. Krishnamurthy, D.-H. Kim, A. Celedon, G. D. Longmore, D. Wirtz(2010, May). A distinctive role for focal adhesion proteins in three-dimensional cell motility. Nat Cell Biol. 2010 Jun;12(6):598-604. doi: 10.1038/ncb2062. 
\bibitem{J. H.-C. Wang}
J. H.-C. Wang, J.-S. Lin.(2007, January). Cell traction force and measurement methods. Biomech Model Mechanobiol 6, 361-371 (2007). https://doi.org/10.1007/s10237-006-0068-4.
\bibitem{Gardel}
M. L. Gardel, B. Sabass, L. Ji, G. Danuser, U. S. Schwarz, C. M. Waterman. (2008, December). Traction stress in focal adhesions correlates biphasically with actin retrograde flow speed. J Cell Biol. 2008 Dec 15;183(6):999-1005. doi:10.1083/jcb.200810060.
\bibitem{Case}
L. B. Case, C. M. Waterman. (2015, August). Integration of actin dynamics and cell adhesion by a three-dimensional, mechanosensitive molecular clutch. Nat Cell Biol. 2015 Aug;17(8):955-63. doi: 10.1038/ncb3191. 
\bibitem{Cramer}
L. P. Cramer. (2013, June). Mechanism of cell rear retraction in migrating cells. Curr Opin Cell Biol. 2013 Oct;25(5):591-9. doi: 10.1016/j.ceb.2013.05.001.
  \bibitem{Bershadsky}
A. D. Bershadsky, M. M. Kozlov. (2011, December). Crawling cell locomotion revisited. Proc Natl Acad Sci U S A. 2011 Dec 20;108(51):20275-6. doi: 10.1073/pnas.1116814108.
  \bibitem{Safsten}
C. A. Safsten, V. Rybalko, L. Berlyand (2022, February). Asymptotic stability of contraction-driven cell motion. Phys Rev E. 2022 Feb;105(2-1):024403. doi: 10.1103/PhysRevE.105.024403. 

  \bibitem{Putelat} 
  T. Putelat, P. Recho, L. Truskinovsky(2018, January). Mechanical stress as a regulator of cell motility. Phys Rev E. 2018 Jan;97(1-1):012410. doi: 10.1103/PhysRevE.97.012410. 
  \bibitem{Recho}
  P. Recho, T. Putelat, L. Truskinovsky. (2015, November). Mechanics of motility initiation and motility arrest in crawling cells. Journal of the Mechanics and Physics of Solids, 84, 469-505. doi: 10.1016/j.jmps.2015.08.006.

  \bibitem{Vabishevich}
  A. Samarskiy, P. Vabishevich. Computational heat transfer. URSS. Moscow; 2003. Available from: http://samarskii.ru/
  \bibitem{Golosnoy}
  I. Golosnoy, J. Sykulski. Evaluation of the front-fixing method capabilities for numerical modelling of diffusion in moving systems. 2008. IET 7th International Conference on Computation in Electromagnetics, Brighton, 2008, pp. 94-95, doi: 10.1049/cp:20080228.
\bibitem{Samarskiy}
  A. Tikhonov, A, Samarskiy. (1962). Homogeneous difference schemes on uneven grids. Journal of Computational Mathematics and Mathematical Physics, 2(5), pp. 812-832. doi: 10.1016/0041-5553(63)90148-4.
\bibitem{Runge-Kutta}
  U. Ascher, L. Petzold. (1998, January). Computer Methods for Ordinary Differential Equations and Differential-Algebraic Equations. Philadelphia: Society for Industrial and Applied Mathematics. doi:10.1137/1.9781611971392.
\bibitem{Barnhart}
Barnhart, E. et al. (2015, April). Balance between cell-substrate adhesion and myosin contraction determines the frequency of motility initiation in fish keratocytes, Proceedings of the National Academy of Sciences, 112(16), pp. 5045-5050, doi:10.1073/pnas.1417257112. 
\end{thebibliography}
\end{document}